\documentclass[12pt,showpacs,amsmath,aps,prc]{revtex4}
\usepackage{amsfonts} 
\usepackage{mathrsfs}
\usepackage{subfigure}
\usepackage{supertabular}
\usepackage{color,graphicx}
\usepackage{multirow}
\usepackage[bookmarksopen]{hyperref}

\def  \th   {\theta}

\def  \veps {\varepsilon}
\def  \del  {\partial}

\def  \bef  {\begin{figure}}
\def  \eef  {\end{figure}}
\def  \beq  {\begin{eqnarray}}
\def  \eeq  {\end{eqnarray}}
\def  \nn   {\nonumber}
\def  \bm   {\bibitem}

\begin{document}
\title{Correlation corrections to the thermodynamic properties
of spin asymmetric QGP matter}
\author {Kausik Pal}
\affiliation {Department of Physics, Serampore College, 
Serampore 712201, India.}

\medskip

\begin{abstract}
We calculate the free energy, entropy and pressure of 
the Quark Gluon Plasma (QGP) at finite temperature and density with 
a given fraction of spin-up and spin-down quarks using a MIT bag model 
with corrections up to ${\cal O} (g^4 \ln g^2)$. The expressions
for the specific heat and the spin susceptibility are derived
in terms of Fermi momentum and temperature. The effects of interaction
between the quarks on the properties of the QGP phase are also investigated.
Within our phenomenological model, we estimate the transition temperature 
$T_c$ by constructing the phase boundary between the hadronic phase and 
the QGP phase. Finally, we compute the equation of state of the QGP 
and its dependence on the temperature and the density.

\end{abstract}
\vspace{0.08 cm}


\maketitle

\section{Introduction}

One of the active areas of high energy physics research is the study
of thermodynamic properties of interacting hadronic matter 
in the extreme conditions of temperature and/or density for the last 
few decades. At such high temperature and/or density
the hadrons are expected to dissolve into their more fundamental
constituents {\it viz.} quarks and gluons, forming a new state
of matter called Quark Gluon Plasma (QGP) \cite{hwa1,hwa2,hwa3}.
The basic theory that describes the strong interaction in terms of 
quarks and gluons is known as Quantum Chromodynamics (QCD). 
It is known that QCD is asymptotically free {\it i.e.} strong 
interaction becomes weak for processes involving high transferred momenta 
\cite{gross,gross73,polit73}.
This means that the quarks and gluons appear to be weakly coupled
at very short distances. At large separations, the effective coupling
becomes progressively stronger resulting in the phenomenon
called the confinement of color charges {\it i.e.} at low density
and temperature, the quarks and gluons remain confined in color
singlet hadrons that constitute hadronic or nuclear matter
\cite{hooft78,hooft81,wilson74}. However,
when the density or the temperature are high enough, the quarks and
gluons start to play dominant roles in determining the thermodynamic
properties of the system. At extreme temperature and/or density 
one expects a transition of hadronic matter into a phase dominated
by quarks and gluons, where color is deconfined and the interaction 
becomes screened 
\cite{collins75,kisslinger76,shuryak80,shuryak84,alford08,schmitt04,blaizot}. 
Such a phase, where any length scale of interest is greater than the screening length of the interaction, 
is known as QGP. At vanishing chemical potential, lattice 
QCD suggests that this phase transition happens at a 
temperature around $T_c\simeq 170-190$ MeV \cite{aoki09,cheng06}. 
At non-zero chemical potential, predictions on the order and 
exact location of the phase transition differ among the various 
lattice groups \cite{fodor02,fodor03,kitazawa02,abuki02,nishida05}.
This is because lattice quantities are not properly defined in such 
a case so that different groups investigated with different additional
approximations.

The possibility of creating high temperature QGP by colliding heavy ions
in the laboratory and studying this phase of matter has been the goal
of experiments at CERN SPS and at the Relativistic Heavy Ion Collider
(RHIC) facility at Brookhaven National Laboratory (BNL)
\cite{NA50_00,STAR94,PHENIX94}. ALICE, ATLAS
and CMS Collaborations at the Large Hadron Collider (LHC) 
have provided further impetus to these studies
\cite{ALICE08,ATLAS10,CMS11}. This experimental
search of the QGP needs reliable theoretical estimates of various
signals which depend on the pressure, entropy, deconfinement
temperature
and the equation of state (EOS) etc. \cite{mod08,mod13,gardim07,plumari12}.

In recent years, significant progress has been
made to understand the behavior of QGP phase of matter leading
to major advancement in the theoretical front addressing some
of the subtle issues of the quasiparticles excitation in
such an environment 
\cite{plumari12,mod13,mod08,bordbar13,rebhan02,gardim07,srivas12}.
One of the major developments, in this context, has been the Hard 
Thermal Loop (HTL) approximated perturbation theory where
these issues are handled in a systematic manner and meaningful
results are obtained after performing suitable resummations 
\cite{blaizot_rept02,rebhan02}. These apart, many calculations
have been performed to study the high temperature transport phenomenon
of QGP including calculations of various transport coefficients
like viscosity, conductivity etc. 
\cite{plumari12,kovtun05,chen13,amato13}. 
Several attempts have also been made to study the EOS for 
interacting QCD matter 
\cite{mod08,mod13,bordbar13,borsanyi10,borsanyi11,srivas12}. 
For example, new lattice results for the equation of state of QCD
with $2+1$ dynamical flavors were obtained in \cite{borsanyi10,borsanyi11}.
Ref.\cite{srivas12} deals with hybrid model in
constructing a deconfining phase boundary between the hadron gas and 
the QGP and provides a realistic EOS for the strongly interacting matter.
Furthermore, in \cite{mod08,mod13} the author have studied the thermodynamic 
properties of weakly interacting unpolarized QGP matter by using a MIT bag model within one gluon exchange (OGE) interaction.
Such investigations have also been performed in 
\cite{bordbar13}, where the calculations have been extended to spin 
polarized matter. 

Being motivated by this series of works, we undertake the present
investigation to study the thermodynamic behavior of weakly interacting 
spin asymmetric QGP matter including correlation corrections by using a 
MIT bag model with non-zero chemical potential. Due to asymptotic
freedom, one may expect that the quarks and gluons interact weakly. 
Thus, the properties of QCD might be computable
perturbatively \cite{kajantie03}. So in our calculations it is assumed 
QCD coupling constant $\alpha_c =\frac{g^2}{4\pi}< 1$. 
It is to be mentioned, that for
weak coupling constant $g$, perturbation theory can only be worked out
to a finite order. In the strong coupling limit, the perturbation
theory fails and one has to resort to lattice results. However,
in the present work we assume the coupling to be weak enough for
the series expansion in terms of $\alpha_c$ to converge.
Although the perturbative expansion converge very slowly, the approach makes
predictions consistent, when comparison is possible, with lattice results 
which do not contain such an approximation \cite{kajantie03}.
In our model for the computation of the thermodynamic quantities, 
we require the knowledge
of total energy density of spin polarized matter with the inclusion
of bubble diagrams \cite{pal_gse}, like what one does for the 
calculation of the correlation energy for degenerate electron 
gas \cite{gellmann57,pines58}. Without such corrections, however,
the calculations are known to remain incomplete as the higher order
terms beyond the exchange diagrams are affected by infrared divergences
due to the exchange of massless gluons \cite{pal_gse}. This indicates
the failure of the naive perturbation series. We know that this problem
can be cured by summing a class of diagrams that makes the perturbation
series convergent and receives logarithmic corrections 
\cite{gellmann57,pines58,pal_gse}. In the present work, 
we calculate various thermodynamic quantities like energy density, 
pressure, entropy density with correlation corrections within the one gluon
exchange model for non-zero chemical potential and show 
how these quantities can be expressed in terms of the spin polarization
parameter $\xi$ and the temperature $T$. We compare some of our 
results with the previous calculations, whenever possible. In addition,
the present work is extended further to estimate the 
specific heat, spin susceptibility of QGP with a given
fraction of spin-up and spin-down quarks. Within our model, we estimate 
the critical temperature $T_c$ for the transition between the hadronic phase
to the quark phase.
Furthermore, we determine the equation of state of QGP and its dependence
on the temperature and the density.

The plan of the article is as follows. In Sec. II, we calculate the 
various thermodynamic quantities with corrections up to 
${\cal O} (g^4 \ln g^2)$. The results are also incorporated in this section. 
Sec. III. is devoted to the summary and discussions.

\section{Thermodynamic properties with correlation}

In this section we calculate the thermodynamic properties,
like energy density $(E)$, pressure $({\cal P})$, 
entropy density $({\cal S})$, free energy $({\cal E})$, 
specific heat $(C_v)$, spin susceptibility $(\chi)$ for QGP matter
with explicit spin dependent quarks by using a MIT bag model with fixed bag 
pressure ${\cal B}$. The MIT bag model considers free particles confined to
a bounded region by the bag pressure. This pressure depends on the 
quark-quark interaction. For short, the bag constant (${\cal B}$) is 
the difference between the energy densities of non-interacting 
and interacting quarks \cite{bordbar13,chodos74}. Within this
model, for the calculation of energy
density and other related quantities, we assume the QGP is  
composed of the light quarks only, {\rm i.e.} the up and down
quarks which interact weakly, and the gluons which are treated as almost
free \cite{mod13}. We consider the color symmetric forward scattering
amplitude of two quarks around the Fermi surface by the one 
gluon exchange interaction. The direct term does not contribute
as it involves the trace of single color matrices, which vanishes, 
while the leading contribution comes from the exchange term \cite{pal09}.
We are dealing with quasiparticles whose spins are all eigenstates
of the spin along a given direction, viz. $z$. The quasiparticle
interaction $(f_{pp'}^{ss'})$ can be decomposed into two parts, 
spin may be either parallel $(s=s')$ or antiparallel $(s=-s')$
corresponding to spin nonflip $(f_{pp'}^{\rm{nf}})$ or spin flip 
$(f_{pp'}^{\rm{f}})$ scattering amplitudes \cite{pal09,tatsumi00}, such that
\beq\label{scat_amp}
f_{pp'}^{ss'} &=& f_{pp'}^{\rm nf}+\frac{1}{2}f_{pp'}^{\rm f},
\eeq  
where $p$ and $p'$ are the momentum of the quasiparticles. Here,
the factor $1/2$ is due to the equal scattering possibilities involving
spin-up spin-down and spin-down spin-up quarks, {\rm i.e.}
$f_{pp'}^{+-}=f_{pp'}^{-+}$, where super-scripts denote 
the spin indices.
 
Since spin and momentum have no preferred direction, we take the 
average over the angles $\th_1$ and $\th_2$ corresponding to spins $s$
and $s'$. The angular averaged interaction parameter is given by
\cite{pal09,pal_sus}
\beq\label{int01}
f_{pp'}^{ss'}&=&\frac{g^2}{9pp'}
\int \frac{\rm d\Omega_{1}}{4\pi}\int \frac{\rm d\Omega_{2}}{4\pi}
\Big[1+({\hat p}\cdot {\hat s})({\hat p'}\cdot {\hat s'})\Big],
\eeq
where $g$ is the coupling constant. 

In our calculations, it is assumed that the nuclear matter is
the initial state with equal neutron and proton densities 
{\rm i.e.} $n_{\rm n}=n_{\rm p}$, so that $n_{\rm u}=n_{\rm d}$, 
which means the contributions of ${\rm u}$ and ${\rm d}$ quarks are equal.
Since, we are dealing with ultra-relativistic massless quarks, all 
of the thermodynamic quantities, as we shall see, are obtained 
as a function of temperature $(T)$ and Fermi momentum $(p_f)$. 
Although the Fermi momentum is not a suitable experimental quantity,  
in fact it is better to calculate baryon density as a function of
temperature and Fermi momentum. 
Since the baryon number of a quark is $\frac{1}{3}$, we have
baryon density $n_b=\frac{1}{3}n_q$ and the quark density $n_q$ is 
given by
\beq\label{qrk_den}
n_q = \sum_{s=\pm} n_q^s
&=& N_c N_f \sum_{s=\pm}\int\frac{d^3p}{(2\pi)^3}n_{p}^s(T)\nn\\
&=& \frac{1}{\pi^2}\sum_{s=\pm}\Big[{p_f^s}^3+\pi^2 T^2 p_f^s\Big].
\eeq
Here, $N_c=3$ and $N_f=2$ are the color and flavor degeneracy factors,
and $n_{p}^{s}(T)$ is the Fermi distribution function. 
Using the above Eq.(\ref{qrk_den}), 
the Fermi momentum as a function of $n_b$ and $T$
can be written as
\beq\label{fermi_mom}
p_f(n_b,T) &=& \Big\{\frac{3}{4}\pi^2 n_b
+\sqrt{\frac{\pi^6 T^6}{216}[(1+{\xi})^{1/3}+(1-{\xi})^{1/3}]^3
+\frac{9}{16}\pi^4 n_b^2}\Big\}^{1/3}\nn\\
&& +\Big\{\frac{3}{4}\pi^2 n_b
-\sqrt{\frac{\pi^6 T^6}{216}[(1+{\xi})^{1/3}+(1-{\xi})^{1/3}]^3
+\frac{9}{16}\pi^4 n_b^2}\Big\}^{1/3}.
\eeq
Here, $\xi=(n_q^{+}-n_q^{-})/(n_q^{+}+n_q^{-})$ 
is the spin polarization parameter with 
$0\le \xi \le 1$. $n_q^+$ and $n_q^-$ correspond to the densities 
of spin-up and spin-down quarks, respectively.

\subsection{Energy, pressure and entropy density of the QGP}

To calculate the total energy density of the QGP, we need to compute
the energy densities for both quark and gluon separately, as we treat
gluons as non-interacting. The leading contributions to the energy density
of quarks are given by three terms viz. kinetic, exchange and 
correlation energy densities, {\rm i.e.}
\beq
E_{q} &=& E_{kin}+E_{ex}+E_{corr}.
\eeq
The total kinetic energy density for spin-up and spin-down quarks,
including the color and flavor degeneracy factors 
for quarks, is 
\beq\label{kin_eng}
E_{kin} &=& \frac{3}{(2\pi)^2}
\Big\{p_f^4[(1+{\xi})^{4/3}+(1-{\xi})^{4/3}]\nn\\
&&+2\pi^2T^2p_f^2[(1+{\xi})^{2/3}+(1-{\xi})^{2/3}]
+\frac{14}{15}\pi^4T^4\Big\},
\eeq 
where, $p_f$ is the Fermi momentum of the unpolarized matter 
$(\xi = 0)$.

For spin asymmetric quarks, the exchange energy
density consists of two terms ${E_{ex}=E_{ex}^{\rm nf}+E_{ex}^{\rm f}}$, 
and can be determined by evaluating the following integrals 
\beq
E_{ex}^{\rm nf}&=&\frac{N_f N_c^2}{2}\sum_{s=\pm}\int\int\frac{d^3p}{(2\pi)^3}
\frac{d^3p'}{(2\pi)^3}f_{pp'}^{\rm nf}~n_{p}^s(T)~n_{p'}^s(T),
\label{exeng_nf}\\
E_{ex}^{\rm f}&=&N_f N_c^2\int\int\frac{d^3p}{(2\pi)^3}
\frac{d^3p'}{(2\pi)^3}f_{pp'}^{\rm f}~n_{p}^s(T)~n_{p'}^s(T).
\label{exeng_f}
\eeq
The analytical expression for the total exchange energy density 
is found to be
\beq\label{ex_eng}
E_{ex}&=&\frac{g^2}{(2\pi)^4}\Big\{p_f^4[(1+{\xi})^{4/3}+
(1-{\xi})^{4/3}+2(1-{\xi}^2)^{2/3}]\nn\\
&&+\frac{4}{3}\pi^2T^2p_f^2[(1+{\xi})^{2/3}+(1-{\xi})^{2/3}] 
+\frac{4}{9}\pi^4T^4\Big\}.
\eeq
It might be noted that the kinetic and the exchange energy density
at low temperature have been calculated in Ref.\cite{pal_sus}, but 
here, the relevant quantities have been derived by retaining
higher-order terms in $T$.

The next higher order correction to the energy density beyond the
exchange term is the correlation energy. The detailed calculation 
of correlation energy for spin polarized matter have been given in 
\cite{pal_gse} which we quote here:

\beq\label{cor_eng1}
E_{corr} & \simeq & \frac{1}{(2\pi)^3}\frac{1}{2}
\int_0^{\pi/2}\sin^2\theta_E {\rm d}\theta_E
\left\{\Pi_L^2\left[\ln\left(\frac{\Pi_L}{\veps_f^2}\right)
-\frac{1}{2}\right]
+2\Pi_T^2\left[\ln\left(\frac{\Pi_T}{\veps_f^2}\right)
-\frac{1}{2}\right]\right\},
\eeq
with $\theta_E=\tan^{-1}(|k|/k_0)$ and $K \equiv (k_{0},0,0,|k|)$ 
is the gluon $4$-momentum. Including the leading finite
temperature corrections to the gluon self-energy, the relevant
$\Pi_L$ and $\Pi_T$ are determined to be 
\cite{pal_gse, andreas_th,blaiz_rept}
\beq
\Pi_L &=&\Big[\frac{N_fg^2}{4\pi^2}\sum_{s=\pm}{p_f^s}^2
+\Big(N_c+\frac{N_f}{2}\Big)\frac{g^2T^2}{3}\Big]
\sin^{-2}\theta_E(1-\theta_E \cot\theta_E),
\label{piL}\\
\Pi_T &=& \Big[\frac{N_fg^2}{4\pi^2}\sum_{s=\pm}{p_f^s}^2
+\Big(N_c+\frac{N_f}{2}\Big)\frac{g^2T^2}{3}\Big]
\cdot \frac{1}{2}[1-\sin^{-2}\theta_E (1-\theta_E \cot\theta_E)].
\label{piT}
\eeq

These are then inserted in Eq.(\ref{cor_eng1}) and performing the
$\theta_E$ integration, we estimate $E_{corr}$. 
The leading $g^4\ln g^2$ order contribution is given by
\beq\label{cor_eng2}
E_{corr} &=& \frac{g^4\ln g^2}{(2\pi)^6}\cdot \frac{1}{8}
\Big\{p_f^4[(1+{\xi})^{4/3}+
(1-{\xi})^{4/3}+2(1-{\xi}^2)^{2/3}]\nn\\
&&+\frac{16}{3}\pi^2T^2p_f^2[(1+{\xi})^{2/3}+(1-{\xi})^{2/3}]
+\frac{64}{9}\pi^4T^4\Big\}.
\eeq

It is to be noted that the exchange energy and the correlation energy 
for various $\xi$ are always positive for massless quarks 
\cite{pal09,mod13,tatsumi00}. Thus, the interaction between the quarks
inside the bag is repulsive and it helps the interacting quarks and the
gluons to go out from the bag very easily. Therefore a transition to a QGP
phase is more favored than in a non-interacting case \cite{mod13}.

For the energy density of gluons we have
\beq\label{eng_gluon}
E_{g} &=& 16 \int\frac{d^3k}{(2\pi)^3}\frac{k}{e^{k/T}-1}\nn\\
&=& \frac{8}{15}\pi^2T^4,
\eeq
where $16$ is the degeneracy of gluons.

Since our system is ultra-relativistic, there is a simple relation between
the pressure and the energy density:
\beq
{\cal P} &=& \frac{1}{3}E.
\eeq

To obtain the total energy density of the QGP, the bag pressure 
$({\cal B})$ should be included in addition to the quark and 
gluon contributions \cite{mod13},
\beq\label{E_QGP}
E_{QGP} &=& E_{q}+E_{g}+{\cal B}.
\eeq
The pressure of the system is determined to be
\beq\label{P_QGP}
{\cal P}_{QGP} &=& {\cal P}_{q}+{\cal P}_{g}-{\cal B}.
\eeq
Once the value of ${\cal P}$ is determined, one can readily calculate
the entropy density of the system by evaluating
\beq
{\cal S}_{QGP} &=& \frac{\del}{\del T}({\cal P}_{QGP}).
\eeq
The thermodynamic properties of the system can be obtained by 
using the Helmholtz free energy relation
\beq
{\cal E}_{QGP} &=& E_{QGP}-T{\cal S}_{QGP}.
\eeq

For the numerical estimation of all these quantities, following 
Refs.{\cite{mod13,mod08}}, 
we take $\alpha_c = \frac{g^2}{4\pi} = 0.2$, as the coupling
constant of QCD and the bag pressure 
${\cal B} = 208$ MeV ${\rm fm^{-3}}$
for zero hadronic pressure.

\vskip 0.22in
\begin{figure}[htb]
\begin{center}
\resizebox{7.5cm}{5.7cm}{\includegraphics[]{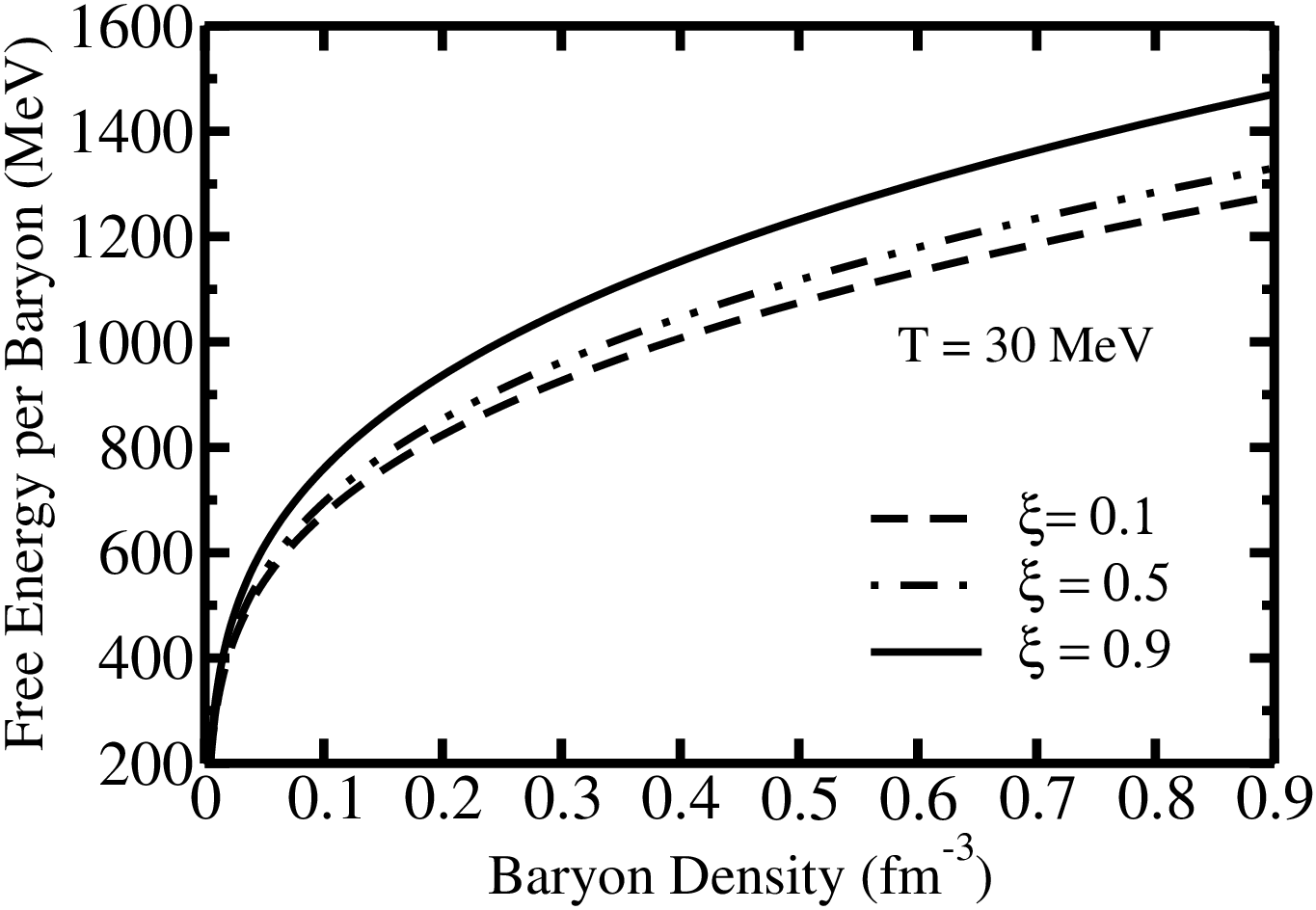}}~~~~
\resizebox{7.5cm}{5.7cm}{\includegraphics[]{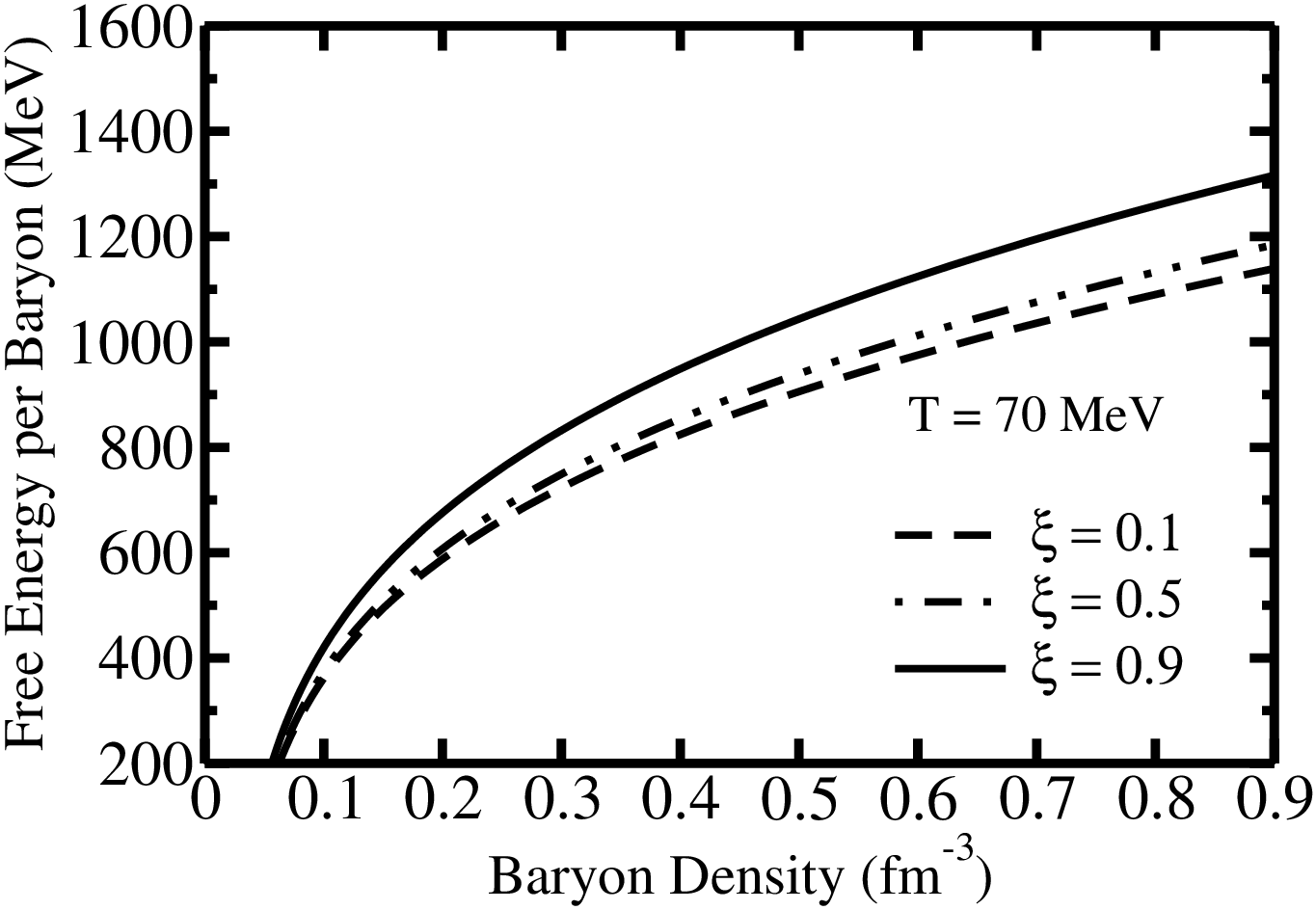}}
\caption{Density dependence of the free energy per baryon at two 
different temperatures for different values of spin polarization $\xi$.}
\label{freng_den}
\end{center}
\end{figure}


In Fig.~{\ref{freng_den}}, we plot the free energy per baryon as a 
function of baryon density for various order parameter $\xi$ at 
two different temperatures, $30$ MeV and $70$ MeV.
This shows that the free energy is larger with higher value of $\xi$.
Therefore QGP is more stable when quarks are unpolarized.

\begin{figure}[htb]
\begin{center}
\resizebox{7.5cm}{5.7cm}{\includegraphics[]{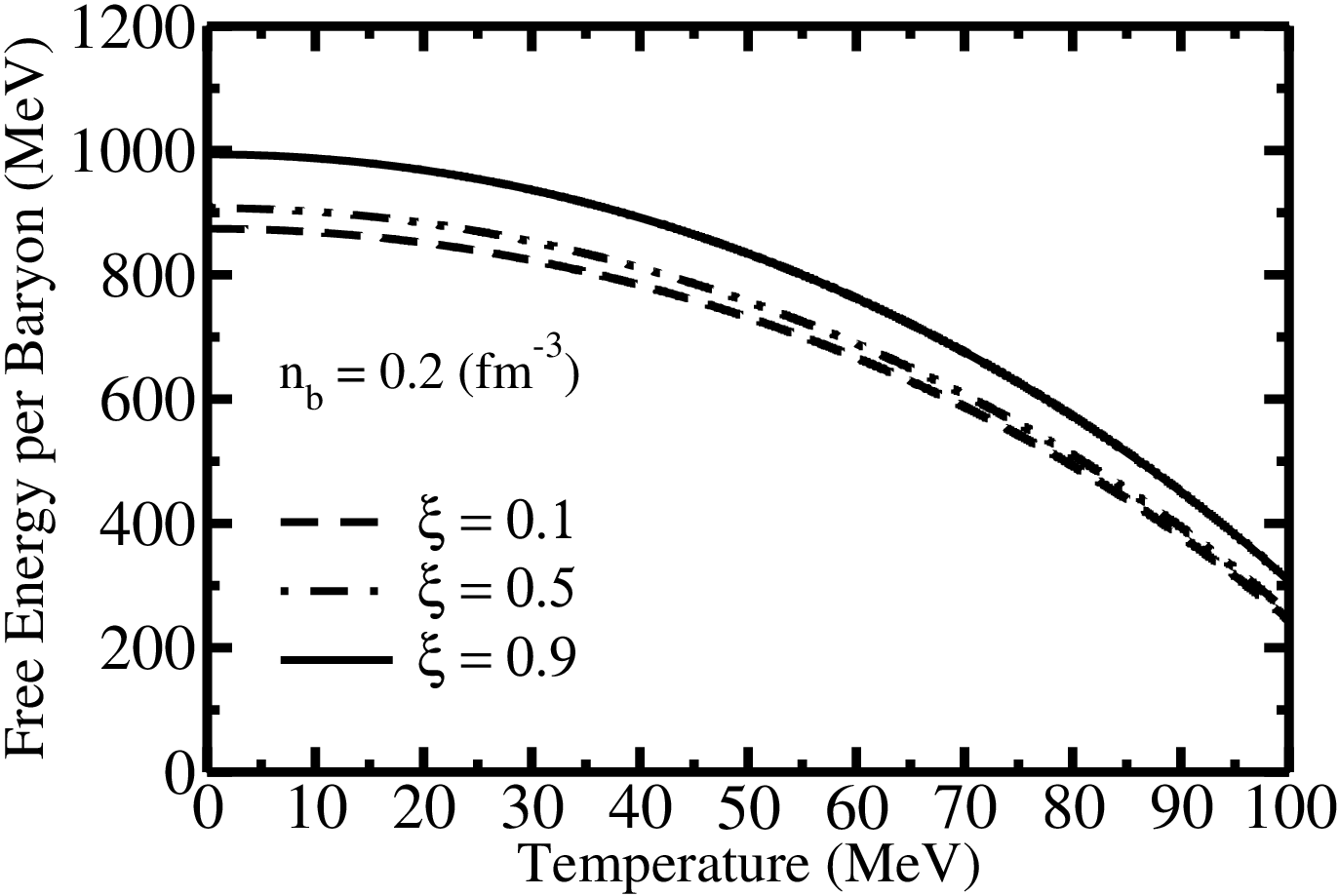}}~~~~
\resizebox{7.5cm}{5.7cm}{\includegraphics[]{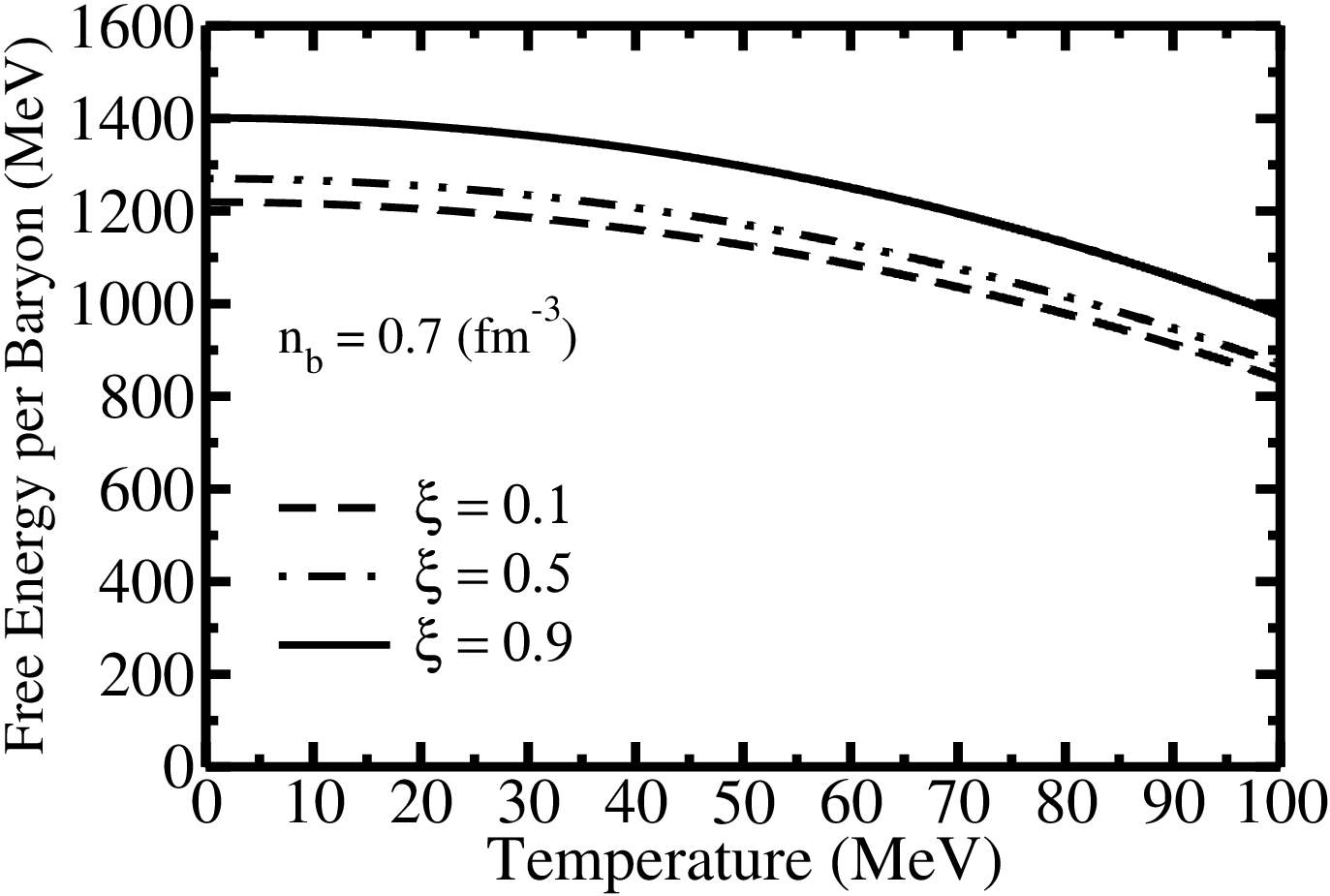}}
\caption{Temperature dependence of the free energy per baryon at two 
different baryon densities for various $\xi$.}
\label{freng_temp}
\end{center}
\end{figure}


Similarly, in Fig.~{\ref{freng_temp}}, we show the temperature dependence
of the free energy with various $\xi$ at two different densities,
$0.2$ ${\rm fm^{-3}}$ and $0.7$ ${\rm fm^{-3}}$ respectively. 
As expected the free energy decreases, as observed both
in Fig.~{\ref{freng_den}} and Fig.~{\ref{freng_temp}}, with increasing
temperature proving that QGP becomes more stable with unpolarized quarks.

\vskip 0.2in
\begin{figure}[htb]
\begin{center}
\resizebox{7.2cm}{5.3cm}{\includegraphics[]{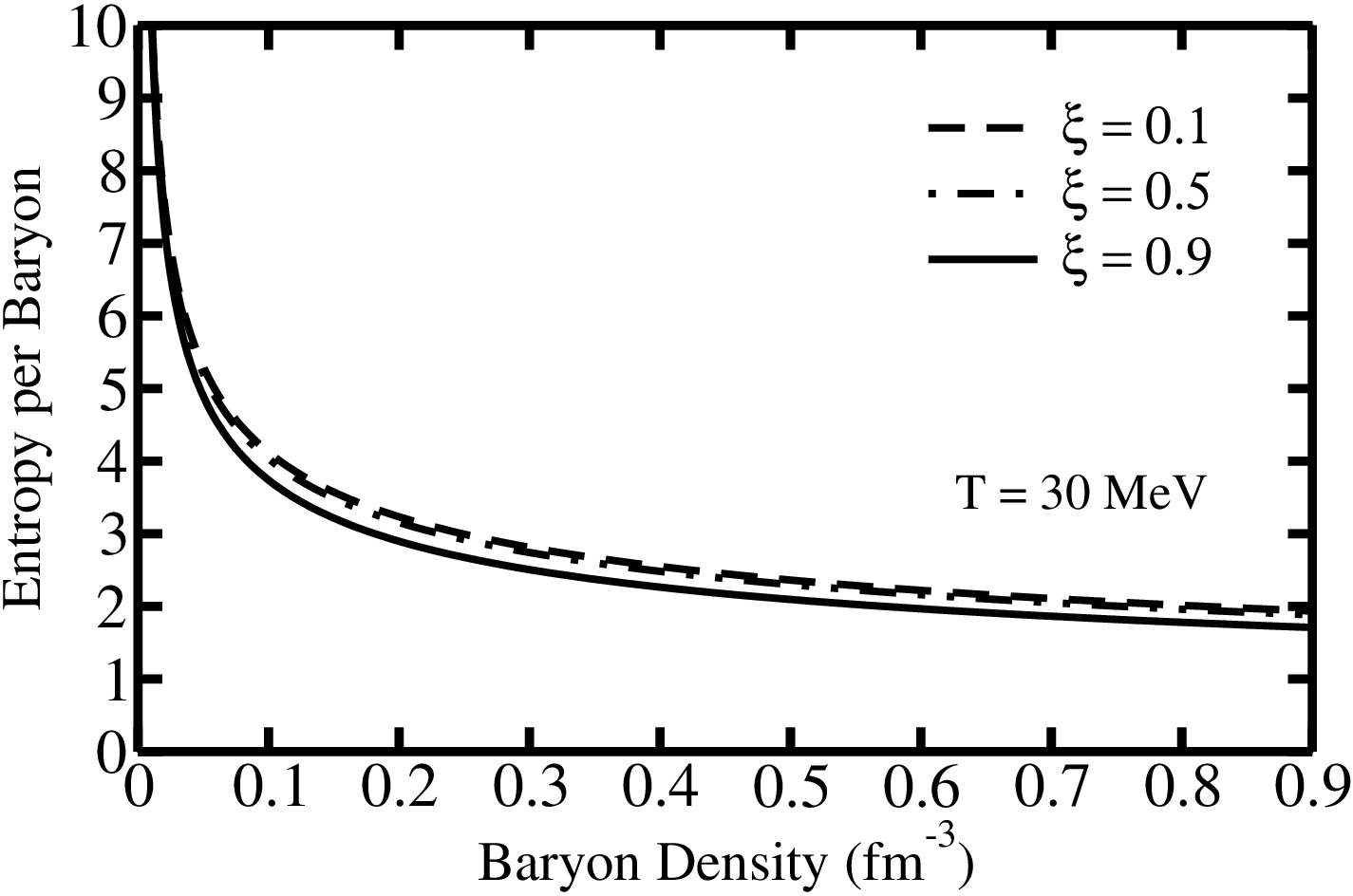}}~~~~~~~~
\resizebox{7.2cm}{5.3cm}{\includegraphics[]{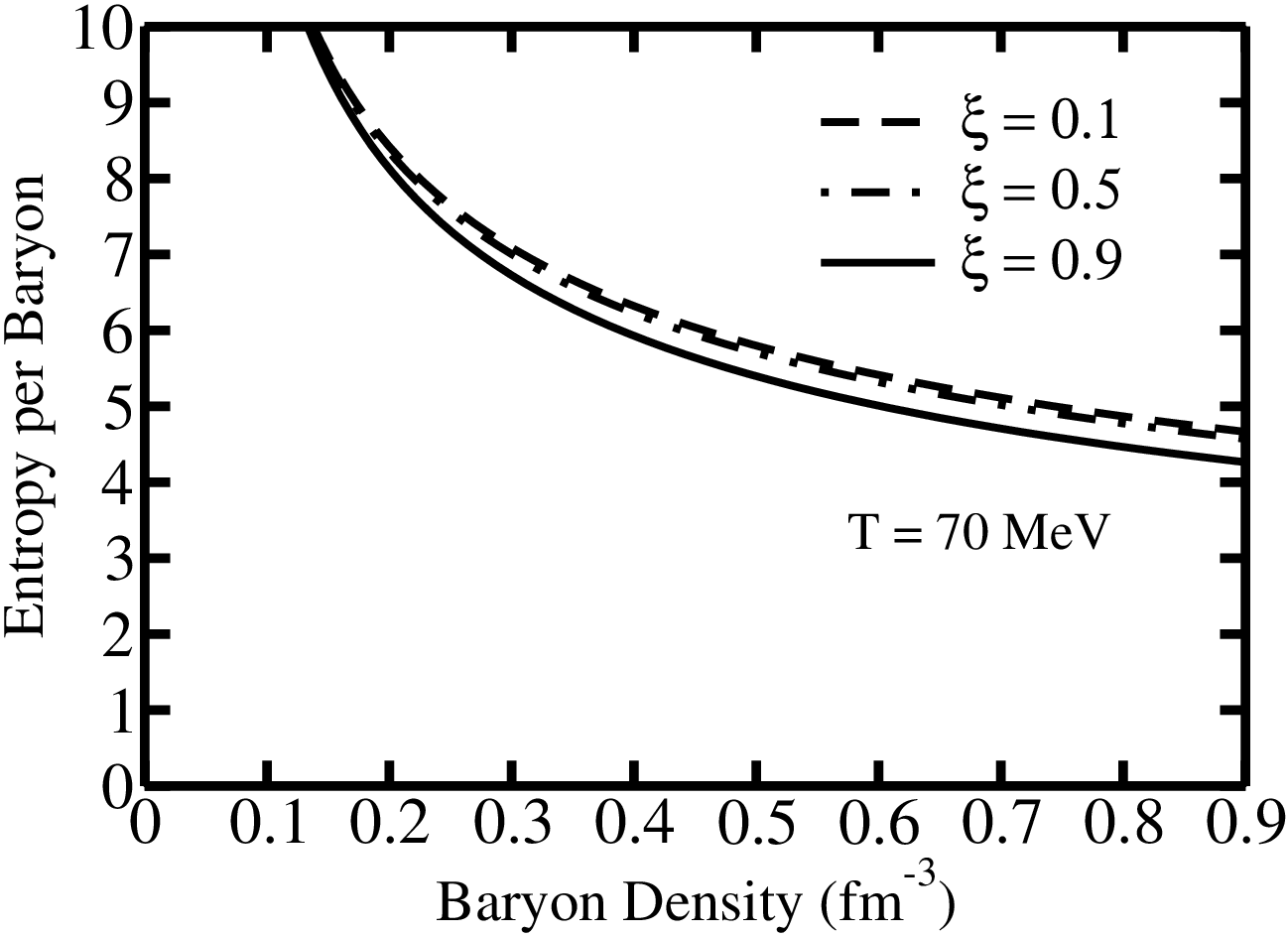}}
\caption{Density dependence of the entropy per baryon at two 
different temperatures for various $\xi$.}
\label{entrp_den}
\end{center}
\end{figure}

\begin{figure}[htb]
\begin{center}
\resizebox{7.2cm}{5.3cm}{\includegraphics[]{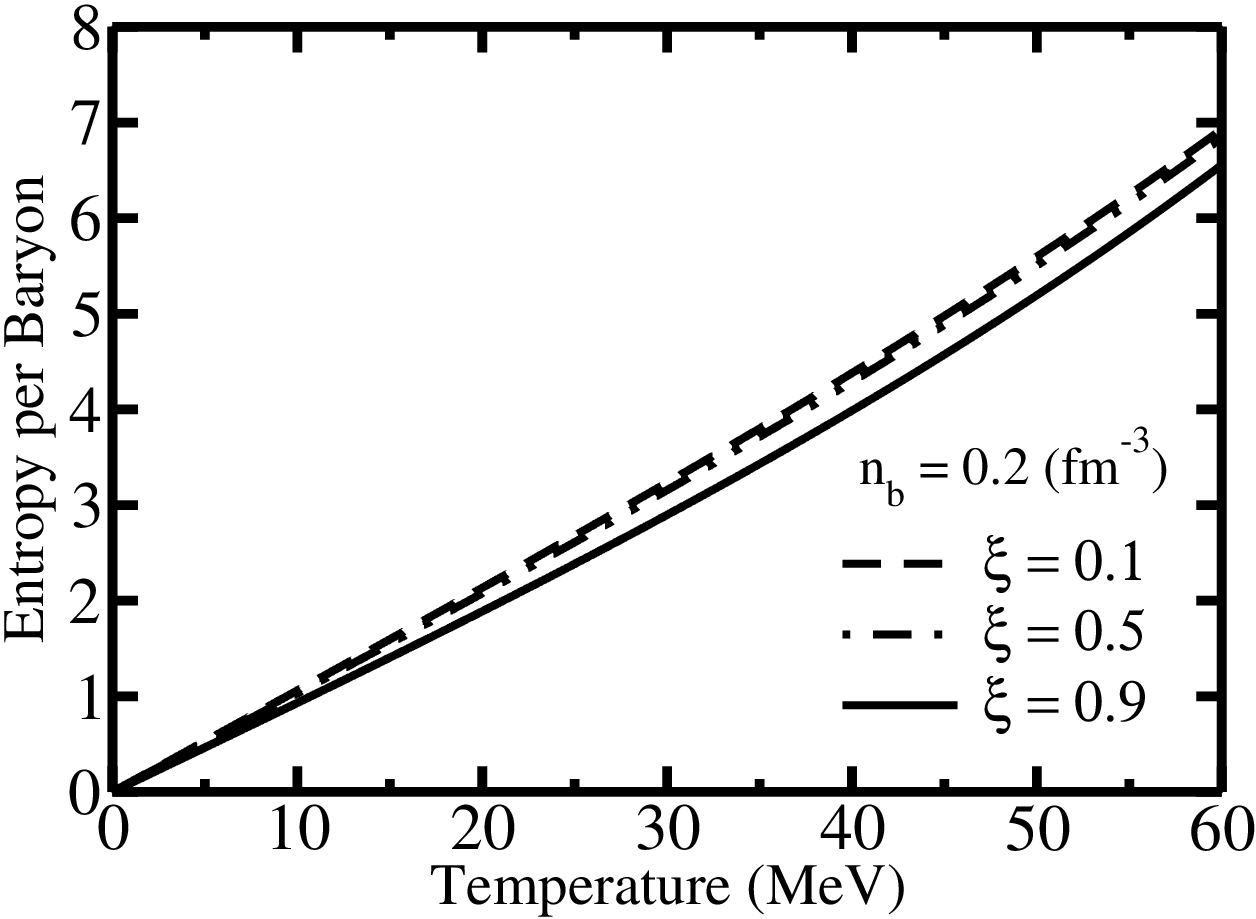}}~~~~~~~~~~
\resizebox{7.2cm}{5.3cm}{\includegraphics[]{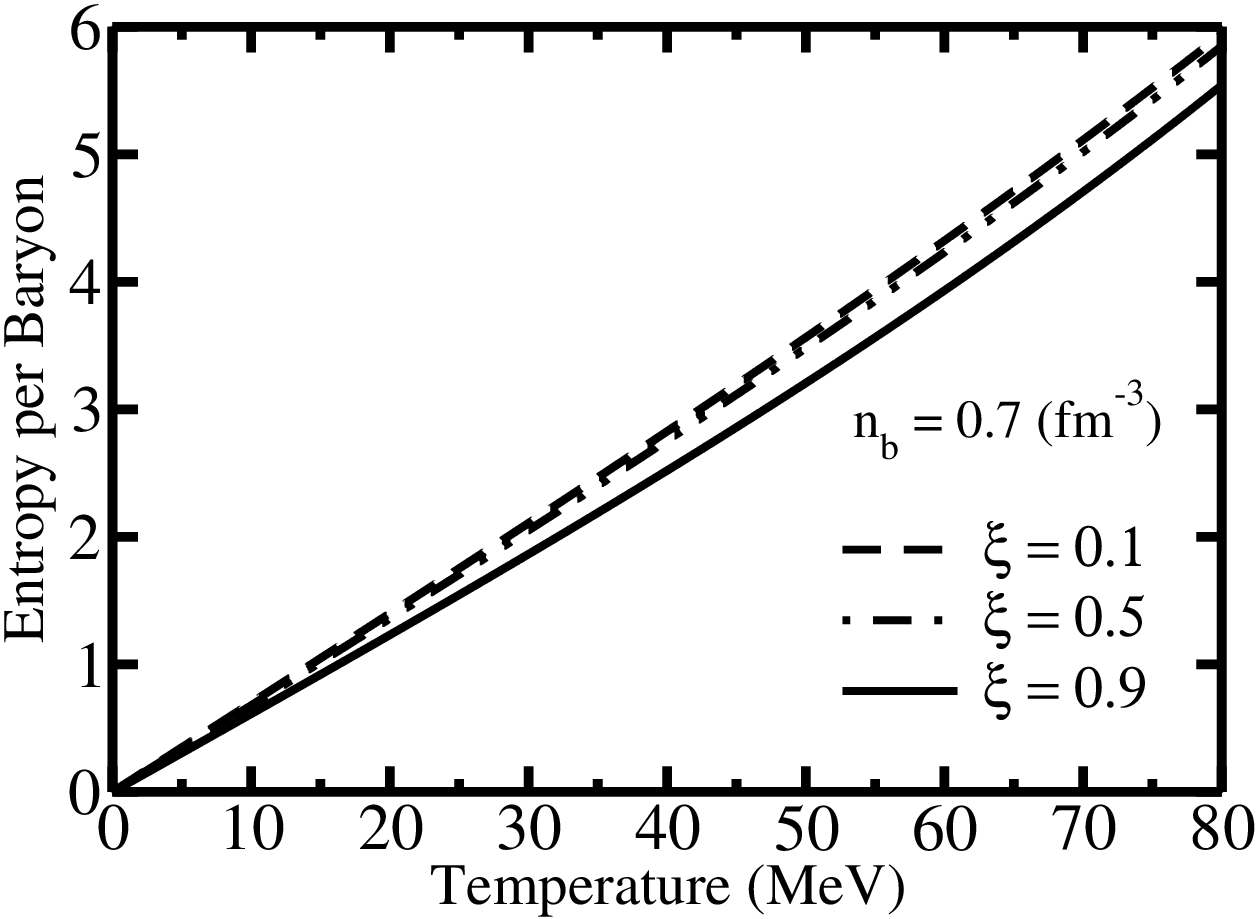}}
\caption{Temperature dependence of the entropy per baryon at two 
different baryon densities for various $\xi$.}
\label{entrp_temp}
\end{center}
\end{figure}

In Fig.~{\ref{entrp_den}}, the variations of entropy per baryon in terms
of baryon density have been plotted with different values of order 
parameter $\xi$. The entropy per baryon is an important quantity connected 
to experimental observables of quark-gluon plasma formation in heavy ion 
collisions \cite{leoni94}. We find that the entropy decreases with 
increasing baryon density and at low density, entropy is approximately
the same for all the values of the order parameter at a fixed temperature.
Similarly, Fig.~{\ref{entrp_temp}} shows the temperature dependence
of the entropy per baryon for various $\xi$. This shows that the 
entropy per baryon in the QGP is an increasing function of 
temperature \cite{mod13} and that the entropy decreases with 
increasing $\xi$. 
The numerical estimates suggest that the entropy per baryon is continuous
along the phase boundary as observed in the right panel 
of Fig.~{\ref{entrp_temp}}, while for the
left one the temperature reached in these computations is not high enough
to make any conclusion.
It should be mentioned that when the entropy per baryon varies continuously
across the phase boundary, it is a smooth cross over.

\vskip 0.2in


\begin{figure}[htb]
\begin{center}
\resizebox{7.5cm}{5.6cm}{\includegraphics[]{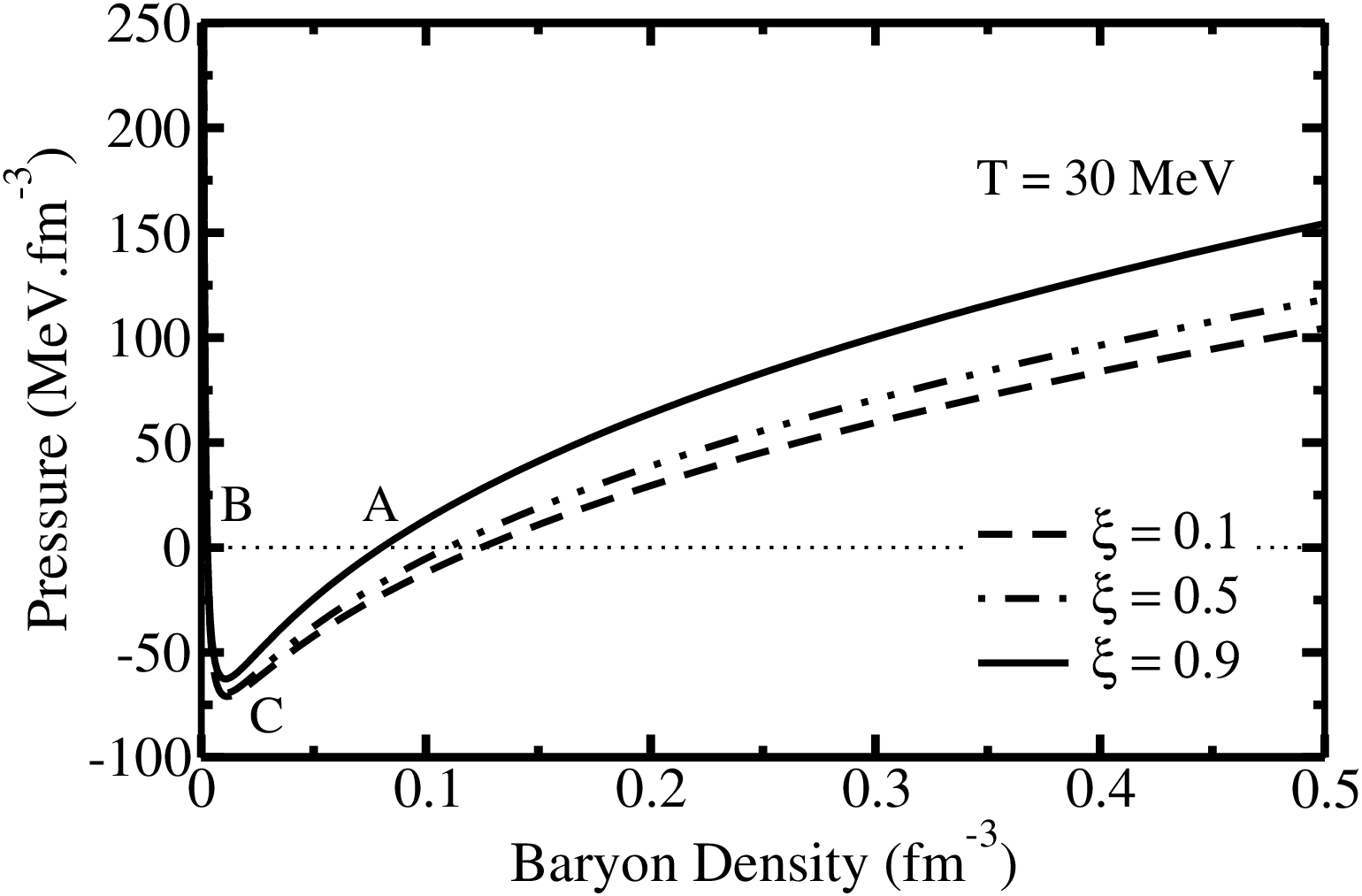}}~~~~~~
\resizebox{7.5cm}{5.6cm}{\includegraphics[]{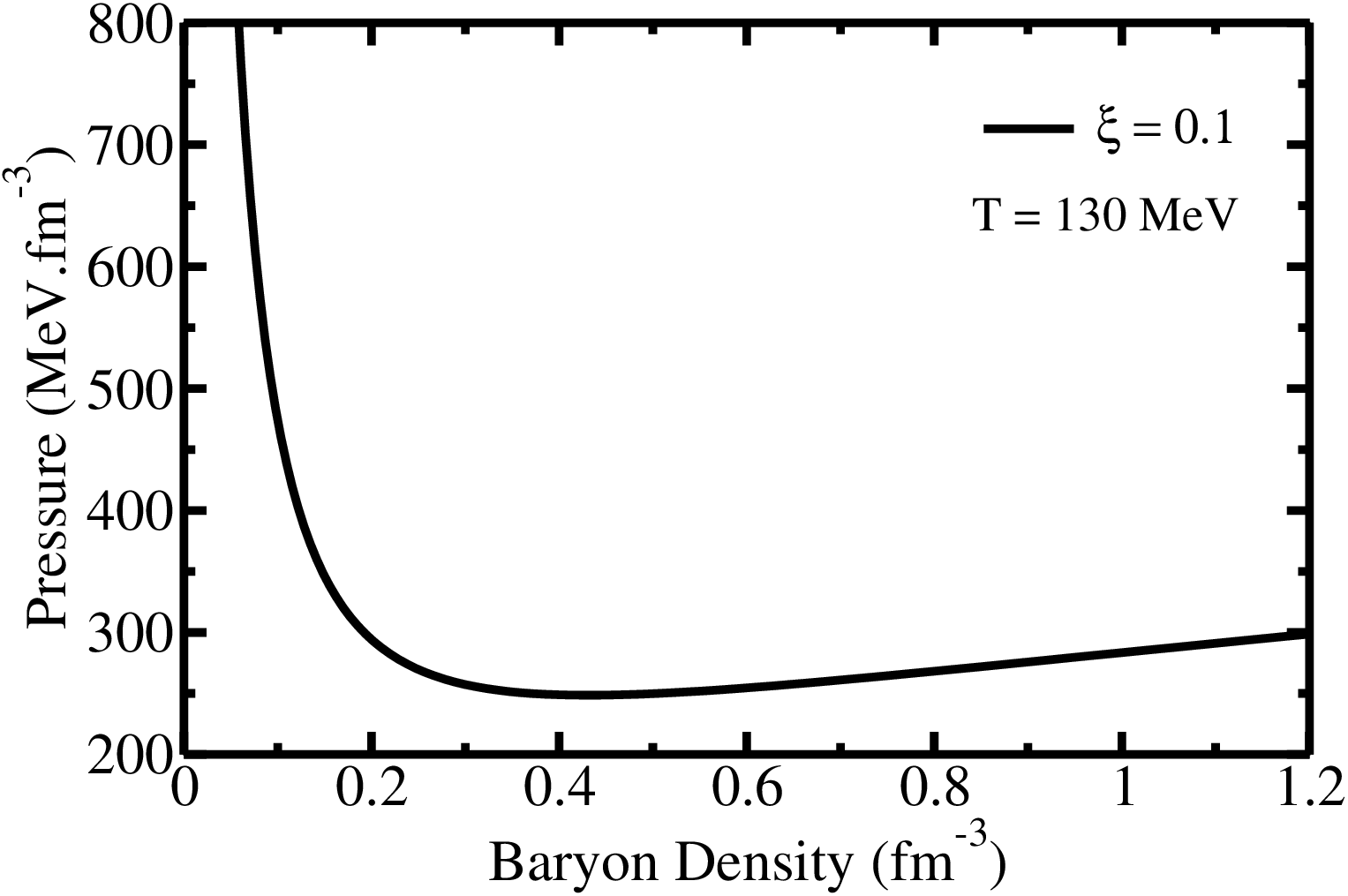}}
\caption{The QGP equation of state as a function of baryon density
for two different temperature with various $\xi$.}
\label{eqn_st01}
\end{center}
\end{figure}


We have also studied the equation of state. The pressure is illustrated
as a function of density in Fig.~{\ref{eqn_st01}} with
$T=30$ MeV (left panel) and $T=130$ MeV (right panel). 
The behavior observed for the pressure is driven by the fact that
the interaction between the particles is attractive at large distance
and repulsive at short distance. 
In the left panel of Fig.~{\ref{eqn_st01}}, the values of pressure 
and baryon density beyond the point `A' correspond to the QGP phase 
while the segment BC correspond to the hadronic phase.
The segment AC denotes the unstable state. 
In the right panel, we observe that
the unstable state shows a tendency to disappear as the temperature
increases. We also see, the pressure increases by increasing the baryon
density and for a fixed density the pressure of QGP is larger with 
polarized quark than the unpolarized one. Therefore, the transition 
to the deconfined phase with polarized quark at lower density would
be favored in comparison to the one with unpolarized quarks.

\vskip 0.22in
\begin{figure}[htb]
\begin{center}
\resizebox{7.5cm}{5.6cm}{\includegraphics[]{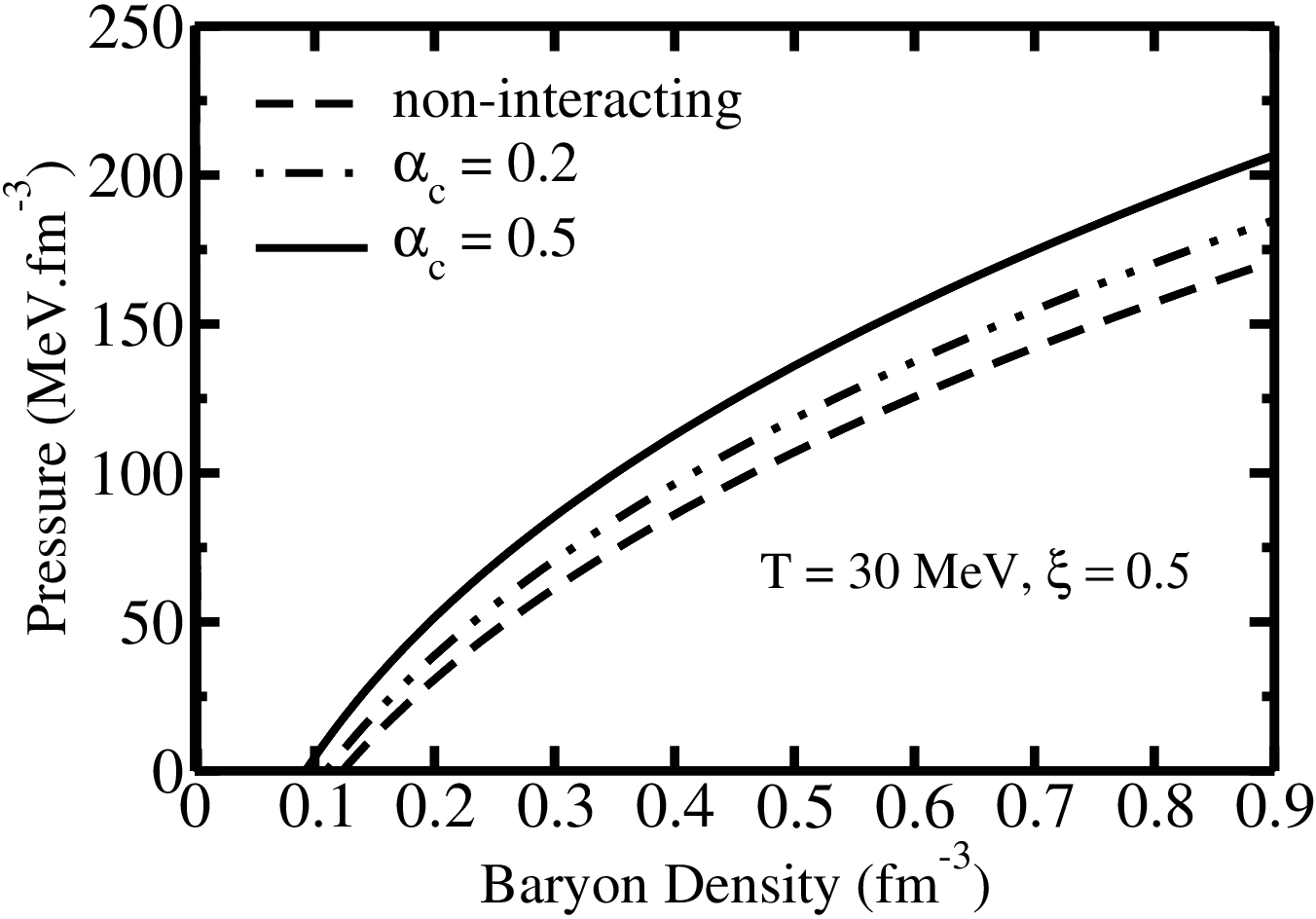}}~~~~~~
\resizebox{7.5cm}{5.6cm}{\includegraphics[]{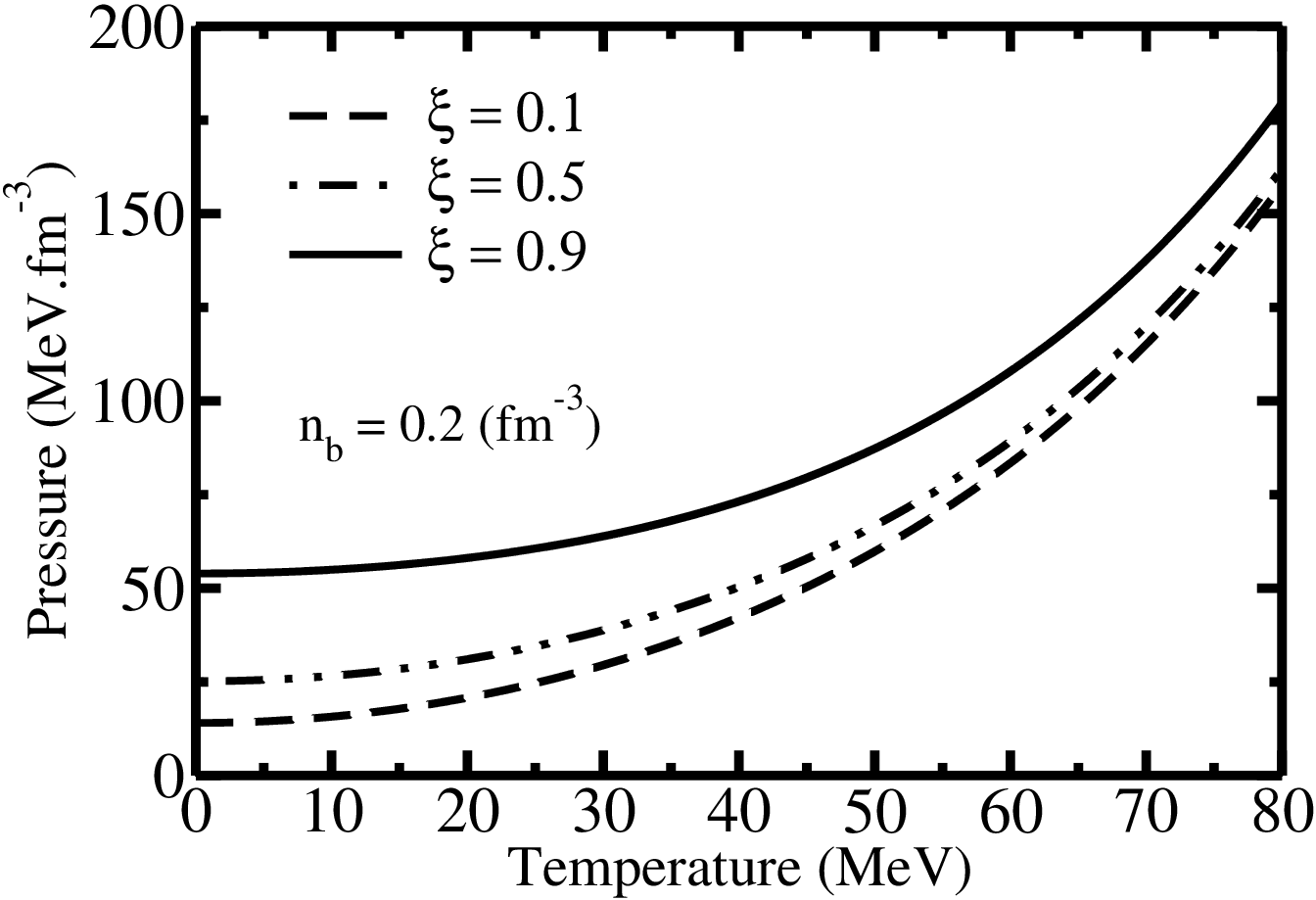}}
\caption{The pressure for different coupling constants
as a function of baryon density (left panel) and the pressure as a function of temperature for various $\xi$ (right panel).}
\label{pres_coup}
\end{center}
\end{figure}


Similarly, in the left panel of Fig.~{\ref{pres_coup}}, the variations of pressure 
of interacting QGP matter with baryon density for two different 
QCD coupling constants, and also for the noninteracting QGP at 
temperature $30$ MeV has been shown. It is shown that 
at the same baryon density the pressure is larger for the interacting cases 
in comparison with the non-interacting one and if the interaction strength
increases also the pressure raises further. Thus the interaction
makes easier the quarks transition to the deconfined phase at lower density
and an increase of the interaction strength pushes the pressure of quarks
to be similar to the bag pressure at smaller baryon density. Therefore, 
the interaction of the QCD coupling constant reduces the 
value of the density to reach a transition \cite{mod13}.

In the right panel of Fig.~{\ref{pres_coup}},
the pressure of spin polarized QGP matter 
as a function of temperature for different $\xi$ has been plotted.
We find that the pressure increases by increasing the temperature. 
It has been observed that at a constant temperature, the pressure of 
QGP is larger for polarized quarks than for unpolarized ones. 
These results indicate that the unpolarized state is energetically 
favorable for the QGP at any temperature and baryon density. The
equation of state of spin asymmetric QGP matter becomes stiffer
by increasing the baryon density (temperature). It has to be noted that 
the pattern is qualitatively similar with that in 
Ref.{\cite{bannur12}}.

\vskip 0.2in
\begin{figure}[htb]
\begin{center}
\resizebox{7.5cm}{5.6cm}{\includegraphics[]{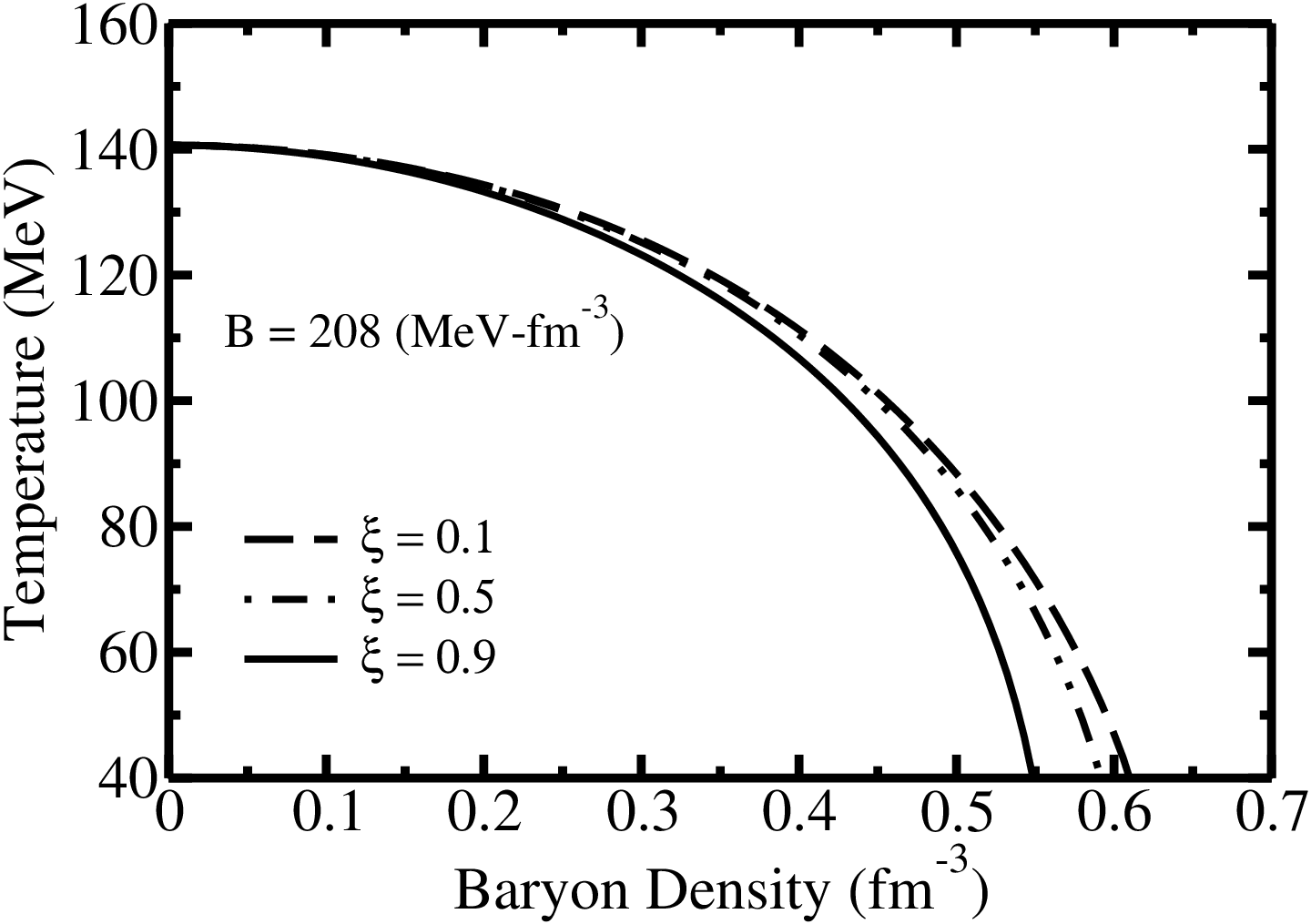}}~~~~~~
\resizebox{7.5cm}{5.6cm}{\includegraphics[]{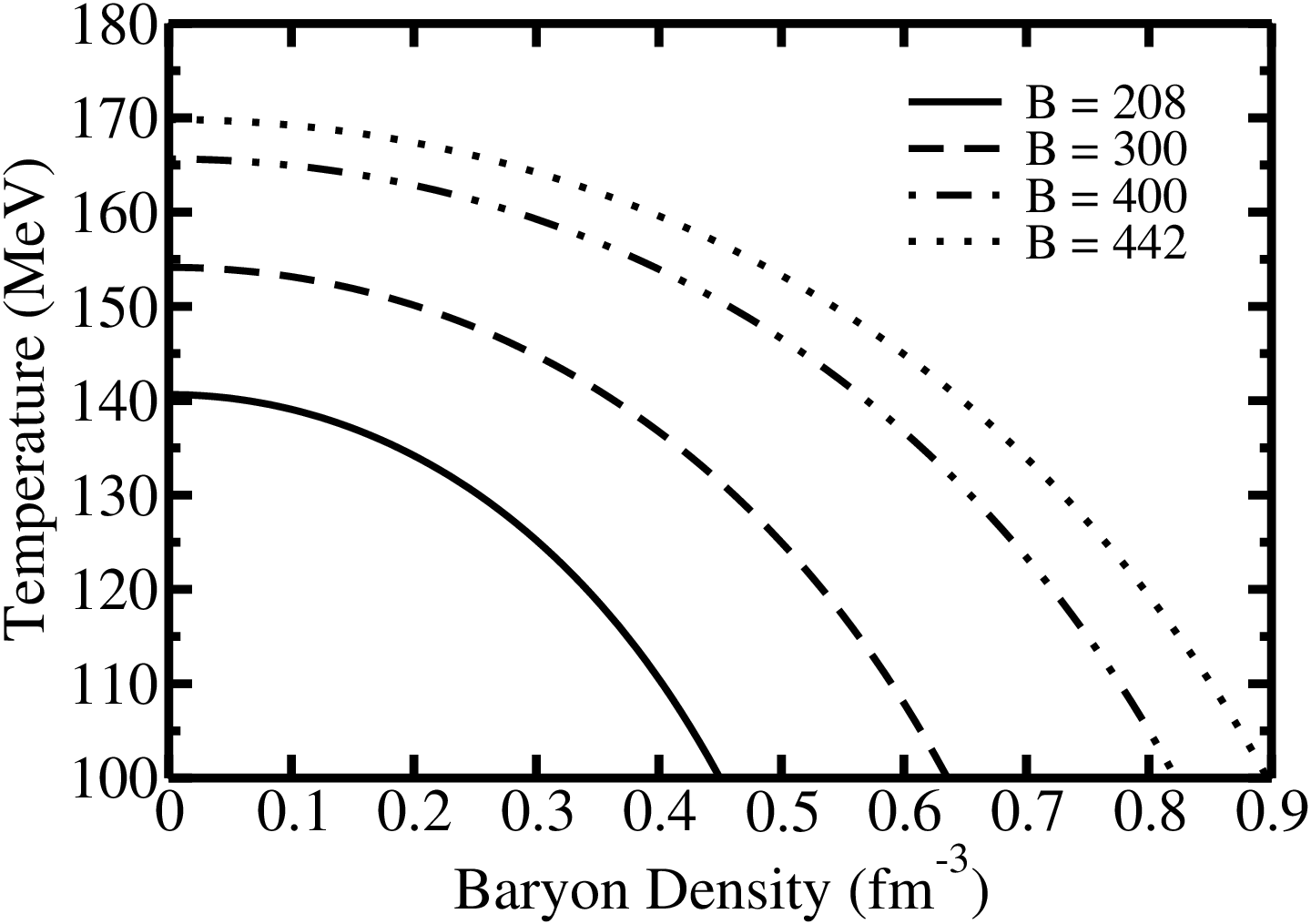}}
\caption{The phase diagram with different $\xi$ (left panel) and
with different bag pressure (right panel).}
\label{ph_dia}
\end{center}
\end{figure}


The QGP phase diagrams are depicted in Fig.~{\ref{ph_dia}};
in the left panel for different $\xi$ at a constant bag pressure 
and in the right panel at different bag pressures for an unpolarized 
QGP. In the left panel, we see that the critical temperature
is independent of the polarization parameter while critical density is
different for unpolarized and polarized QGP. On the other hand, 
in the right panel it is shown that the critical temperature and 
density for the phase transition are different for different 
bag pressures and the critical values increase when increasing the bag 
pressure. This is because, as mentioned earlier, the interaction 
between the quarks is repulsive and helps the quarks to escape from 
the bags. This, in turn, causes the formation of QGP at smaller
baryon densities and temperatures \cite{mod13}. In the right panel
of Fig.~{\ref{ph_dia}}, it is shown that the critical temperature 
for the deconfined
phase transition lies between $130< T_{c}< 170$ MeV. The exact location 
of the phase boundary varies as results are obtained for non-zero 
chemical potential.
When the bag pressure is about $442$ MeV ${\rm fm^{-3}}$, the estimated
results for critical temperature, {\rm i.e.} $T_c = 170$ MeV, is 
consistent with the lattice results 
\cite{QGP05,boyd95,boyd96,karsch02,laermann03,karsch00}.
It has to be mentioned that for our case, the critical parameters are 
determined from the phase boundary by using the condition that the 
bag pressure $(\cal B)$ is independent of chemical potential $(\mu)$ 
and temperature $(T)$, but in realistic cases ${\cal B}$ may depend 
on both.

\subsection{Specific heat}

The specific heat $C_v$ at constant volume is defined as the 
quantity of energy needed to increase the temperature of a system by
one unit, it reads as \cite{landau_book}
\beq
C_v &=& \Big(\frac{\del E_{QGP}}{\del T}\Big)_{v}\nn\\
&=& \Big[1+\frac{g^2}{18\pi^2}+\frac{g^4 \ln g^2}{144\pi^4}\Big]
\cdot \Big[(1+{\xi})^{2/3}+(1-{\xi})^{2/3}\Big]\cdot 3p_f^2T\nn\\
&& +\Big[\frac{74}{15}+\frac{g^2}{9\pi^2}
+\frac{g^4 \ln g^2}{18\pi^4}\Big]\cdot \pi^2 T^3.
\eeq

\vskip 0.2in
\begin{figure}[htb]
\begin{center}
\resizebox{7.5cm}{5.6cm}{\includegraphics[]{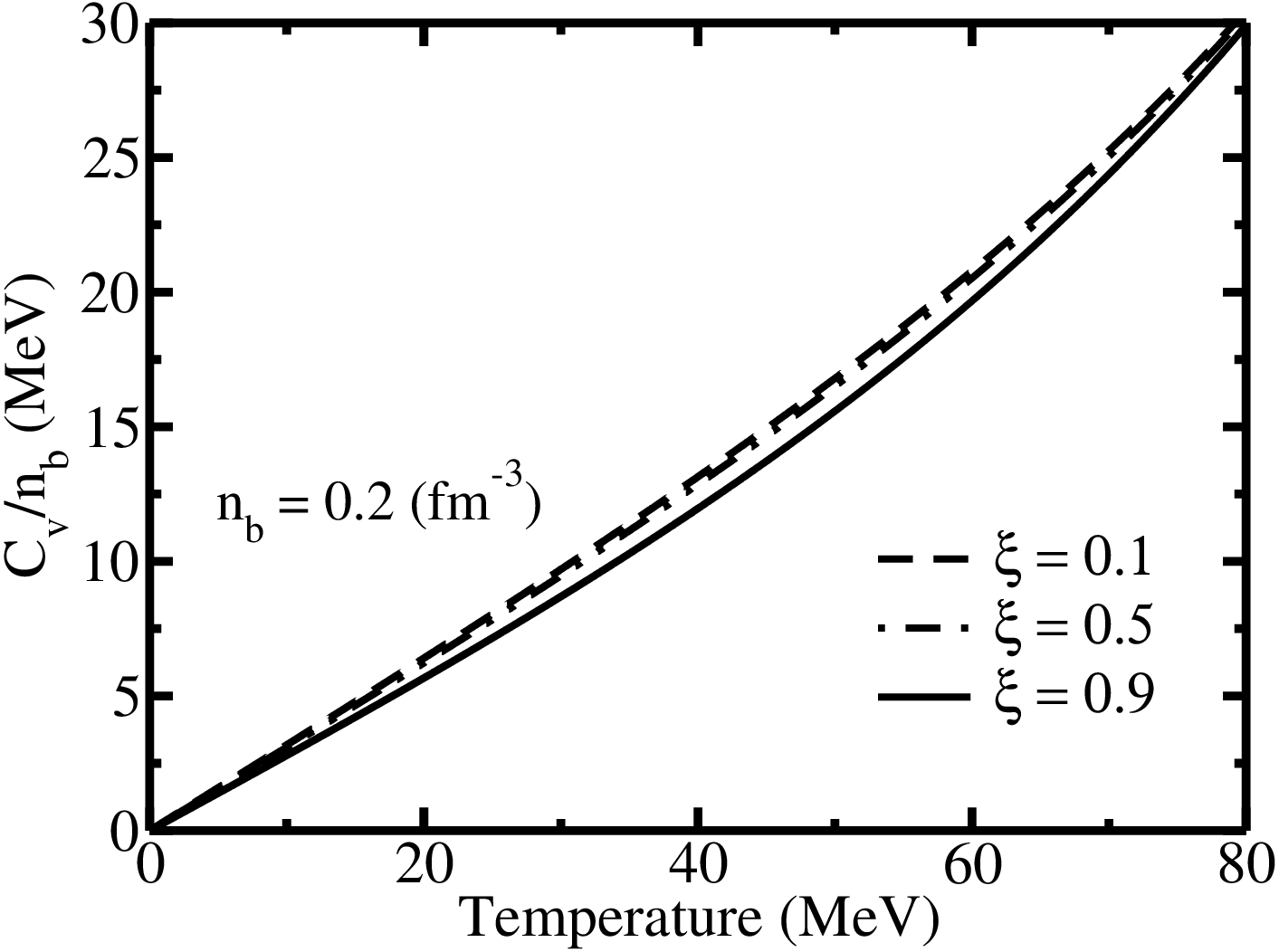}}~~~~~~
\resizebox{7.5cm}{5.6cm}{\includegraphics[]{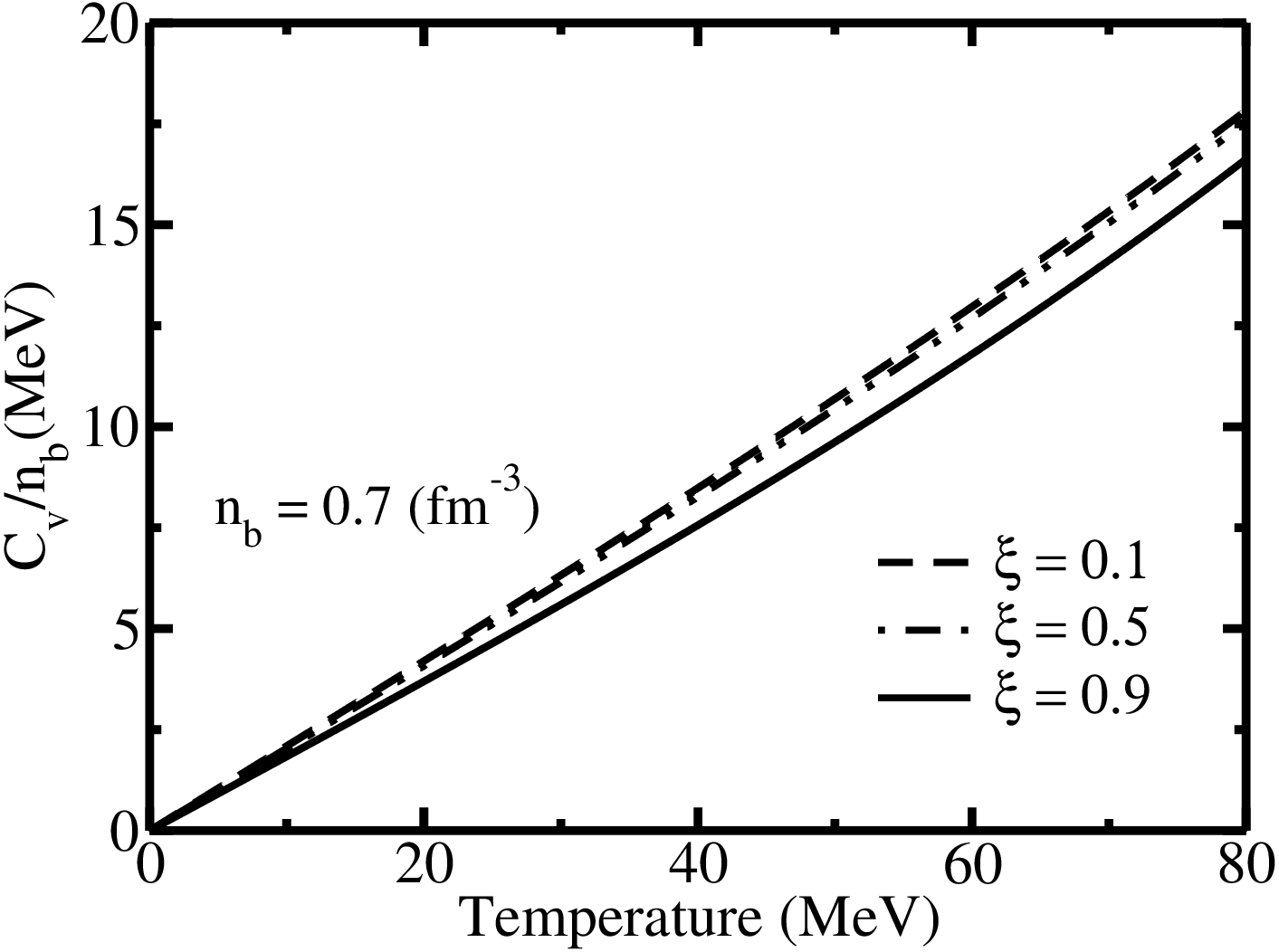}}
\caption{The behavior of specific heat per baryon of QGP at two 
different baryon densities for various $\xi$.}
\label{sp_heat}
\end{center}
\end{figure}


The behavior of specific heat is plotted in Fig.~{\ref{sp_heat}} at
two different densities. We see that the specific heat decreases by 
increasing the baryon density and polarization parameter. This is because
the specific heat is a measure of energy fluctuation with temperature
and the fluctuation is smaller when QGP is polarized. We also observe
that the specific heat indicates complete continuity in its behavior.

\subsection{Spin susceptibility}

The spin susceptibility can be determined by the change in energy of the
system as quarks spin are polarized. In the small $\xi$ limit, the energy
density behaves like 
\cite{pal_sus,perez09,shastry77,shastry78,brueck58,singh89}
\beq\label{xi_expan}
E_{QGP}(\xi)&=& E_{QGP}(\xi=0)+\frac{1}{2}\beta_s\xi^2+{\cal O}(\xi^4),
\eeq
where $\beta_s$ is the spin stiffness constant. The spin susceptibility 
$\chi$ is inversely proportional to the spin stiffness; 
mathematically $\chi=2 \beta_s^{-1}$ \cite{perez09}. Since only quarks are
polarized and the energy density involves three leading terms, the spin susceptibility can be written as \cite{pal_sus}
\beq\label{chi_inv}
\chi^{-1} &=& \chi^{-1}_{kin} +\chi^{-1}_{ex} +\chi^{-1}_{corr}.
\eeq
With the help of Eq.(\ref{xi_expan}), each energy contribution to the 
susceptibility is
\beq\label{sus_cont}
\chi_{kin}^{-1} &=& 
\frac{p_f^4}{3\pi^2}\Big(1-\frac{\pi^2 T^2}{p_f^2}\Big),\nn\\
\chi_{ex}^{-1} &=& 
-\frac{g^2 p_f^4}{18\pi^4}\Big(1-\frac{\pi^2 T^2}{3 p_f^2}\Big),\nn\\
\chi_{corr}^{-1} &=& 
-\frac{g^4 \ln g^2 p_f^4}{576\pi^6}\Big(1+\frac{4\pi^2T^2}{3p_f^2}\Big).
\eeq
Using Eq.(\ref{chi_inv}) and Eq.(\ref{sus_cont}), the sum of 
all the contribution to the susceptibility is given by
\beq\label{spin_sus}
\chi &=& \chi_{P}\Big[1-\frac{g^2}{6\pi^2}
\Big(1+\frac{4\pi^2T^2}{3p_f^2}+\frac{\pi^4T^4}{3p_f^4}\Big)
-\frac{g^4 \ln g^2}{192\pi^4}
\Big(1+\frac{7\pi^2T^2}{3p_f^2}+\frac{4\pi^4T^4}{3p_f^4}\Big)\Big]^{-1}.
\eeq
where $\chi_{P}$ is the Pauli susceptibility \cite{pal_sus,shastry77}.
The value of $\chi$ can be estimated if $g$, $T$ and $p_f$ are exactly
known.

\begin{figure}[htb]
\begin{center}
\resizebox{7.5cm}{5.6cm}{\includegraphics[]{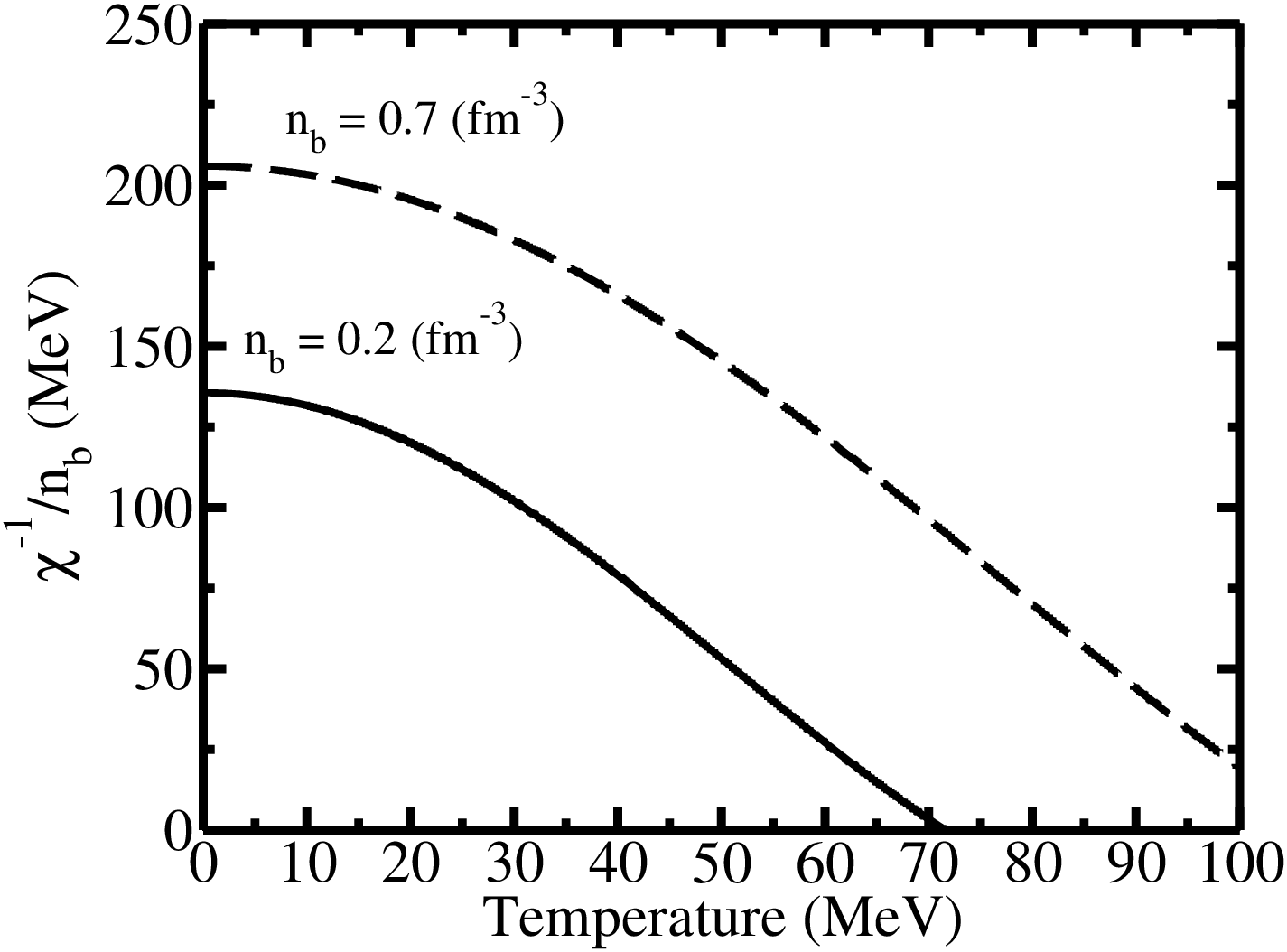}}
\caption{The temperature dependence of inverse spin susceptibility.}
\label{chi}
\end{center}
\end{figure}


The numerical estimation of $\chi^{-1}$ per baryon is given in 
Fig.~{\ref{chi}} for two different densities, $0.2$ ${\rm fm^{-3}}$ 
and $0.7$ ${\rm fm^{-3}}$, respectively. We see susceptibilities
blow up at different temperatures for two different densities.
This can be identified as a physical instability of the QGP matter
towards an ordered phase \cite{shastry77}.

\section{Summary and Discussions}

In this work, we have investigated the thermodynamic properties of the QGP
composed of the massless spin-up and spin-down quarks at
finite temperature and density using a MIT bag model
within one gluon exchange (OGE) interaction. Accordingly, we
calculated the total free energy, entropy density, pressure as a function 
of baryon density and temperature for non-zero chemical 
potential of such a system up to  
${\cal O}(g^4 \ln g^2)$ that includes correction due to correlation
effects. It was shown that the free energy increases by increasing the
polarization parameter. This fact may suggests that the QGP 
with unpolarized quark is energetically favorable. We found that the entropy 
is a decreasing function of the density and an increasing function of the
temperature. In the present phenomenological model, the equation of state 
(EOS) of the QGP with different order parameters has been computed. 
It was shown that the increase of temperature or density makes the EOS 
stiffer. We obtained a phase boundary
for quark-hadron phase transition by making the bag constant independent
of $\mu$ and $T$. The values of critical parameters have been estimated
for the transition from the hadronic phase to the QGP phase by using the EOS.
It has been observed that the entropy per baryon is continuous along the
phase boundary, indicating a cross over from the hadronic phase to the
QGP phase. Moreover, we reported how the OGE interaction 
for the massless quarks affects the phase transition of the QGP and causes 
the system to reach the deconfined phase at smaller baryon densities
and temperatures. In addition, the specific heat and the spin susceptibility
of QGP have also been calculated and the specific heat
decreases by increasing the baryon density in a continuous way.
On the other hand, the spin susceptibility 
blows up at certain temperature for a definite baryon density. This may 
indicate the physical instability of the ordered phase of QGP.

It has to be mentioned that
our results depend on the values of the bag pressure while a change in the 
value of the QCD coupling does not have any dramatic effect in our 
calculations. Moreover, an increase of the bag pressure might improve 
our results towards the lattice QCD calculations. 
Within the scope of the present model, the value of $T_c$ obtained
here, is close to the lattice QCD prediction and the
inclusion of multi-gluon exchange processes may strengthen this conclusion.
More works in this direction are therefore necessary to
examine this issue, especially for multi-flavor systems. 
Leaving aside these questions, our studies might be meaningful for determining the approximate values of temperature, baryon density, 
specific heat etc. regarding the signals of QGP 
and their detection for future studies in the Collider experiments.

\vskip 0.2in
{\bf Acknowledgments}\\

I am indebted to Late Prof. A.K. Dutt-Mazumder who introduced me to 
this topic and his fruitful discussions motivated me to initiate
the present work. I would also like to thank Prof. Jane Alam and 
Prof. Pradip K. Roy for their critical reading of the manuscript.

\end{document}